\documentclass[a4paper,fleqn,usenatbib,useAMS]{mnras}

\usepackage{amsmath}
\usepackage{amssymb}
\usepackage{graphicx}
\usepackage{epstopdf}
\usepackage{inputenc}
\usepackage{float}
\usepackage{geometry}
\usepackage{pdflscape}
\usepackage{bm}

\usepackage[T1]{fontenc}
\usepackage{ae,aecompl}

\usepackage{newtxtext,newtxmath}

\title[Galactic Orbital Effects on Pulsar Timing]{Galactic Orbital Effects on Pulsar Timing}

\author[K. Heflin \& R. Lieu]{K. Heflin\thanks{Contact e-mail: \href{mailto:kth0006@uah.edu}{kth0006@uah.edu}} \& R. Lieu
\\
University of Alabama in Huntsville, Huntsville, AL, USA\thanks{Present address: 301 Sparkman Dr NW, Huntsville, AL 35899}}

\pubyear{2020}

\begin{document}
\label{firstpage}
\pagerange{\pageref{firstpage}--\pageref{lastpage}}
\maketitle

\begin{abstract}
In the currently accepted paradigm, dark matter is hypothesized as an explanation of the flat rotation curves of galaxies under the assumption of virialized orbits. The use of millisecond pulsar timing as a probe of Galactic dark matter content is explored as a means of relaxing this assumption.  A method of inference of the Galactic potential using the frequency derivative $\dot{\nu}$ is produced, and an estimate for a virialized Galactic rotation curve is given through direct observation of acceleration.  The data set used includes 210 pulsars with known $\dot{\nu}$ and astrometric properties, a subset of which also have measured $\ddot{\nu}$. In principle, this enables the exploration of kinematic effects, but in practice, $\ddot{\nu}$ values are found to be too imprecise at present to adequately constrain radial velocities of pulsars. Additionally, surface magnetic field strengths are inferred from $\dot{\nu}$ and the magnetic spin-down contribution to $\ddot{\nu}$ is estimated.  For several pulsars the radial velocity is known, and the kinematic contribution to $\ddot{\nu}$ is estimated accordingly. The binary orbital periods of PSR J1713+0747 and other binary pulsars are also used to constrain Galactic mass density models.
\end{abstract}




\section{Introduction}
Since the discovery of the binary pulsar PSR 1913+16 by \citet{Hulse}, pulsars have been explored as observational tools for a variety of effects both intrinsic and extrinsic to the pulsar system. Millisecond pulsars (MSPs) in particular have afforded high-precision timing, rivaling that of atomic clocks, due primarily to their age and rotational stability.  In this work, we seek to use MSP timing as a tool in constraining Galactic dark matter.

The use of galactic rotation curves, in conjunction with other methods, has been widely regarded as a successful means for inferring dark matter content.  Such an approach is limited, however, by the assumption that a measurement of instantaneous velocity characterizes the acceleration of an object, as is the case for, e.g., a circular orbit. In general, observation of instantaneous velocity is insufficient in ascertaining acceleration and, by extension, in ascertaining force and mass acting on orbiting bodies.

In \S 2, we derive relevant formulae from the equation for Doppler shift, while also introducing several Galactic models to be compared and constrained in this work. We discuss competing effects and observational restrictions on our approach. In \S 3, our method is applied and a summary of the resulting data is given.  \S 4 provides a discussion of our conclusions and possibilities for future work.

\section{Derivation of Relevant Equations}
\subsection{Time-Differentiation of Doppler Shift}
The formulae used in our approach can be derived in a fairly straightforward manner from the equation for Doppler shift,
\begin{equation}\label{Eq1}
\frac{\nu-\nu_0}{\nu}=-\frac{\dot{\textbf{r}}\cdot\textbf{n}}{c},
\end{equation}
where
\begin{equation}\label{Eq2}
\textbf{n}\equiv\frac{\textbf{r}}{|\textbf{r}|},
\end{equation}
$\textbf{r}$ is the vector describing the position of the pulsar with respect to our Sun, $|\textbf{r}|$ its magnitude, $\nu_0$ is the emitted frequency of the pulsar (either the binary orbital frequency or the spin frequency), $\nu$ is the observed frequency, and a dot denotes a time derivative. Since we expect that $(\nu-\nu_0) \ll \nu$, we say
\begin{equation}\label{Eq3}
\frac{\nu-\nu_0}{\nu_0}\approx-\frac{\dot{\textbf{r}}\cdot\textbf{n}}{c}.
\end{equation}
Taking the time derivative,
\begin{equation}\label{Eq4}
\frac{\dot{\nu}}{\nu_0}=-\frac{\ddot{\textbf{r}}\cdot\textbf{n}}{c}-\frac{\dot{\textbf{r}}\cdot\dot{\textbf{n}}}{c},
\end{equation}
where
\begin{equation}\label{Eq5}
\dot{\textbf{n}}=\frac{d}{dt}\left(\frac{\textbf{r}}{|\textbf{r}|}\right)=\frac{\dot{\textbf{r}}}{|\textbf{r}|}-\frac{d|\textbf{r}|}{dt}\frac{\textbf{n}}{|\textbf{r}|}.
\end{equation}
Consider the second term in \eqref{Eq4},
\begin{equation}\label{Eq6}
-\frac{\dot{\textbf{r}}\cdot\dot{\textbf{n}}}{c}=-\frac{\dot{\textbf{r}}^2}{c|\textbf{r}|}+\frac{d|\textbf{r}|}{dt}\frac{(\dot{\textbf{r}}\cdot\textbf{n})}{c|\textbf{r}|}=-\frac{\dot{\textbf{r}}^2}{c|\textbf{r}|}+\frac{d|\textbf{r}|}{dt}\frac{|\dot{\textbf{r}}|}{c|\textbf{r}|}\cos{\theta},
\end{equation}
where $\theta$ is the angle between $\textbf{r}$ and $\dot{\textbf{r}}$. By letting
\begin{equation}\label{Eq7}
\dot{\textbf{r}}=\dot{\textbf{r}}_r + \dot{\textbf{r}}_t=\frac{d|\textbf{r}|}{dt}\textbf{n}+\mu|\textbf{r}|\textbf{t},
\end{equation}
where $\dot{\textbf{r}}_r$ is the line-of-sight velocity vector and $\dot{\textbf{r}}_t$ is the transverse velocity, with magnitude $|\dot{\textbf{r}}_t|=\mu |\textbf{r}|$ and direction $\dot{\textbf{r}}_t\parallel\textbf{t}\perp\textbf{n}$, where $\mu$ is proper motion, we obtain
\begin{equation}\label{Eq8}
\dot{\textbf{r}}^2=\dot{\textbf{r}}_r^2+\dot{\textbf{r}}_t^2=\left(\frac{d|\textbf{r}|}{dt}\right)^2+\mu^2|\textbf{r}|^2.
\end{equation} 
Now, \eqref{Eq4}, \eqref{Eq6}, and \eqref{Eq8} yield
\begin{equation}\label{Eq9}
\frac{\dot{\nu}}{\nu_0}=-\frac{\ddot{\textbf{r}}\cdot{\textbf{n}}}{c}-\frac{\mu^2|\textbf{r}|}{c}-\frac{1}{c|\textbf{r}|}\left(\frac{d|\textbf{r}|}{dt}\right)^2+\frac{d|\textbf{r}|}{dt}\frac{|\dot{\textbf{r}}|}{c|\textbf{r}|}\cos{\theta}.
\end{equation}
Noting that
\begin{equation}\label{Eq10}
\cos{\theta}=\frac{1}{|\dot{\textbf{r}}|}\frac{d|\textbf{r}|}{dt},
\end{equation}
we see that
\begin{equation}\label{Eq11}
\frac{\dot{\nu}}{\nu_0}=-\frac{\ddot{\textbf{r}}\cdot{\textbf{n}}}{c}-\frac{\mu^2|\textbf{r}|}{c}.
\end{equation}
Compare this with the equation obtained by, e.g., \citet{Damour},
\begin{equation}\label{Eq12}
\frac{\dot{P}}{P} = \frac{\textbf{n}_{10}\cdot(\textbf{a}_1-\textbf{a}_0)}{c}+\frac{\mu^2d}{c},
\end{equation}
where $\textbf{a}_1-\textbf{a}_0=\dot{\textbf{r}}$, $\textbf{n}_{10}=\textbf{n}$, and $d=|\textbf{r}|$. We see that since
\begin{equation}\label{Eq13}
\frac{\dot{P}}{P}=-\frac{\dot{\nu}}{\nu},
\end{equation}
\eqref{Eq12} is in agreement with \eqref{Eq11}, where the first term is hereafter referred to as the "radial acceleration" contribution to $\dot{\nu}$, and the second term is the Shklovskii effect.

It is possible to estimate Galactic mass content by assuming the values for $\dot{\nu}/\nu$ are due entirely to acceleration. This means letting
\begin{equation}\label{Accel Term}
\frac{\dot{\nu}}{\nu_0}=-\frac{\ddot{\bm{r}}\cdot\bm{n}}{c},
\end{equation}
where
\begin{equation}
\ddot{\bm{r}}=\ddot{\bm{r}}_p-\ddot{\bm{r}}_\odot=-\nabla\Phi(\bm{r}_p)+\nabla\Phi(\bm{r}_\odot).
\end{equation}
Here, the subscripts $\odot$ and $p$ denote solar and pulsar quantities, respectively, and we must use the $\dot{\nu}/\nu_0$ values after the radial acceleration effect (which has already been corrected for using flat-rotation curves) is replaced.  Since the Shklovskii effect is only dependent on transverse velocity and distance, it will remain unchanged regardless of acceleration.  

Consider the time derivative of equation \eqref{Accel Term}:
\begin{equation}
\frac{\ddot{\nu}}{\nu_0}=-\frac{\dddot{\bm{r}}\cdot\bm{n}}{c}-\frac{\ddot{\bm{r}}\cdot\dot{\bm{n}}}{c}=-\frac{\dddot{\bm{r}}\cdot\bm{n}}{c}-\frac{\ddot{\bm{r}}\cdot\dot{\bm{r}}}{c|\bm{r}|}+\frac{\ddot{\bm{r}}\cdot\bm{r}}{c|\bm{r}|^2}\frac{d|\bm{r}|}{dt},
\end{equation}
or
\begin{equation}\label{Kinematic Contribution}
\frac{\ddot{\nu}}{\nu_0}=-\frac{\dddot{\bm{r}}\cdot\bm{n}}{c}-\frac{\ddot{\bm{r}}\cdot\dot{\bm{r}}}{c|\bm{r}|}+\frac{\ddot{\bm{r}}\cdot\bm{r}}{c|\bm{r}|^2}v_r.
\end{equation}

The absence of measured values for radial velocities of an overwhelming majority of pulsars necessitates the parameterization of this quantity in the above model, after which a best-fit value for $\dot{\textbf{r}}_p$ is obtained through minimization of $\ddot{\nu}-\ddot{\nu}_{\textrm{predicted}}(\dot{\textbf{r}}_p)$ in cases where the transverse velocity of a pulsar is known.

The absence of magnetic spin-down effects in binary orbital period derivatives and the negligibility of orbital decay due to gravitational wave emission in wide binaries would appear to make binary orbital periods more suited to constrain acceleration and, by extension, Galactic mass. We shall see, however, that the scarcity of data and the lack of ephemeral corrections yields a problem of similar severity in binary orbital timing. Ephemeral corrections to PSR J1713+0747 by \citet{Zhu}, however, make this pulsar a better candidate for constraint of Galactic dark matter, as we shall see.

\subsection{Galactic Models}
With the exception of the Galactic model put forth by \citet{Kenyon}, the models of the following paragraphs are presented in the form of a mass density distribution $\rho$, which necessitates the calculation of the Galactic potential $\Phi$. Since
\begin{equation}
\nabla^2\Phi=4\pi G \rho,
\end{equation}
description of the acceleration
\begin{equation}
\bm{a}=-\nabla\Phi
\end{equation}
necessitates solution of Poisson's equation numerically.  We accomplish this by utilizing Green's functions, which is especially computationally demanding when we seek to try many values for parameters of $\rho$ as a match for the observed acceleration effect.

\subsubsection{Kenyon Potential}
A Galactic potential suggested by data obtained from \textit{Gaia} according to \citet{Brown} is that of the three-component model of \citeauthor{Kenyon}. \citeauthor{Kenyon} describe the Galactic potential $\Phi_G$ by decomposition into several terms: $\Phi_b$, which describes the potential of the Galactic bulge; $\Phi_d$, which describes the Galactic disk; and $\Phi_h$, which describes the Galactic halo, i.e.,
\begin{equation}
\Phi_G=\Phi_b+\Phi_d+\Phi_h.
\end{equation}
Each component, in turn, is given by
\begin{equation}
\Phi_b(r)=-\frac{GM_b}{r+a_b},
\end{equation}
where $M_b = 3.76\times10^9M_\odot$, $a_b=0.1 \textrm{ kpc}$,
\begin{equation}
\Phi_d(\rho, z)=-\frac{GM_d}{\sqrt{\rho^2+\left[a_d+\left(z^2+b_d^2\right)^{1/2}\right]^2}},
\end{equation}
where $M_d=6\times 10^{10}M_\odot$, $a_d=2750\textrm{ kpc}$, and $b_d=0.3\textrm{ kpc}$, and
\begin{equation}
\Phi_h(r)=-\frac{GM_h}{r}\ln(1+r/r_h),
\end{equation}
where $M_h=10^{12}M_\odot$, and $r_h=20\textrm{ kpc}$. Together, these yield a flat rotation curve $V(3\textrm{ kpc} < r < 50 \textrm{ kpc})\approx 235\textrm{ km/s}$.  We shall use pulsar timing to constrain $M_h$, though, in principle, any one of the potential components can be constrained using this approach.

The values for acceleration as determined by
\begin{equation}
\ddot{\textbf{r}}_p=-\nabla\Phi_G(\textbf{r}_p)=-\nabla\Phi_b(\textbf{r}_p)-\nabla\Phi_d(\textbf{r}_p)-\nabla\Phi_h(\textbf{r}_p)
\end{equation}
are calculated numerically for each pulsar, with the halo mass $M_h$ parameterized.

In considerations of $\ddot{\nu}/\nu_0$, the time derivative of acceleration (or jerk), given by
\begin{equation}
\dddot{\textbf{r}}_p=-\frac{d[\nabla\Phi_G(\textbf{r}_p)]}{dt}=-\frac{d[\nabla\Phi_b(\textbf{r}_p)]}{dt}-\frac{d[\nabla\Phi_d(\textbf{r}_p)]}{dt}-\frac{d[\nabla\Phi_h(\textbf{r}_p)]}{dt},
\end{equation}
is also calculated numerically, this time with the radial velocity $v_r$ parameterized. While the dependence on $v_r$ of the third term in \eqref{Kinematic Contribution} is obvious, the other two terms also depend on $v_r$, since $\dot{\textbf{r}}=\dot{\textbf{r}}(v_r)$ and $\dddot{\textbf{r}}=\dddot{\textbf{r}}(v_r)$.

\subsubsection{Sofue Model}
\citet{Sofue} proposed a Galactic mass distribution described by four components: two bulge components, one component describing the Galactic disk, and one describing the Galactic dark matter halo.  The bulge components are characterized using
\begin{equation}
\rho_b(R,z)=\frac{\rho_{0,b}}{(1+r/r_0)^\alpha}\exp[-(r/r_{cut})^{\zeta}],
\end{equation}
where $r=\sqrt{R^2+(z/q)^2}$, and $\rho_{0,b}$ and $r_0$ are characteristic densities and radii, respectively. In the Sofue model, $\zeta=1$, $\alpha=0$, $q=1$, and $r_0=1 \textrm{ kpc}$. For the inner bulge, $\rho_{0,b}=3.7\times 10^4 M_\odot/\textrm{pc}^3$, and the cutoff radius $r_{cut}=0.0035\textrm{ kpc}$. The second bulge, referred to as the main bulge, has $\rho_{0,b}=2.1\times 10^2 M_\odot/\textrm{pc}^3$ and $r_{cut}=0.12\textrm{ kpc}$. 

The disk component is described by the function
\begin{equation}
\rho_d(R,z)=\frac{\Sigma_{0,d}}{(2z_d)^{\eta}}\exp\left[-\frac{|z|}{z_d}-\frac{R}{R_d}\right],
\end{equation}
where $\Sigma_{0,d}$ is a characteristic surface density, and $z_d$ and $R_d$ are the scale height and length, respectively. In the Sofue model, a flat disk is used such that $z_d=0$ and $\eta=0$. Additionally, $\Sigma_{0,d}=292 M_\odot/\textrm{pc}^2$ and $R_d=4.0\textrm{ kpc}$.

The Galactic dark matter halo is described according to
\begin{equation}
\rho_h(R,z)=\frac{\rho_{0,h}}{X(1+X)^2},
\end{equation}
where $X=\sqrt{R^2+z^2}/h$, with characteristic density $\rho_{0,h}$ and characteristic radius $h$. Sofue takes $\rho_{0,h}=0.029 M_\odot/\textrm{pc}^3$ and $h=10\textrm{ kpc}$.

Because we are interested in constraint of Galactic dark matter content, $\rho_{0,h}$ and $h$ will be the subject of our test for the Sofue model.

\subsubsection{McMillan Model}
\citet{McMillan} proposed a Galactic model comprised of six components. These components have similar functional forms to those of the Sofue model but with different parameters. \citeauthor{McMillan} used a single bulge with $\zeta=2$,$\alpha=1.8$,$r_0=0.0075\textrm{ kpc}$,$q=0.5$,$\rho_{0,b}=98.4 M_\odot/\textrm{pc}^3$, and $r_{cut}=2.1\textrm{ kpc}$.

\citeauthor{McMillan} employs four Galactic disk components: one thick, one thin, one describing $H_1$ gas, and one describing $H_2$ gas. The thick and thin stellar disks have the same functional form as the disk of \citeauthor{Sofue}, but with $\eta=1$ for both. The disk heights are $z_d=0.3\textrm{ kpc}$ and $z_d=0.9\textrm{ kpc}$ for the thin and thick disks, respectively. For the thin disk, $\Sigma_{0,d}=896 M_\odot/\textrm{pc}^2$ and $R_d=2.5\textrm{ kpc}$.  For the thick disk, $\Sigma_{0,d}=183 M_\odot/\textrm{pc}^2$ and $R_d=3.02\textrm{ kpc}$.

The $H_1$ and $H_2$ disks are described by
\begin{equation}
\rho_g(R,z)=\frac{\Sigma_{0,d}}{4z_d}\exp\left[-\frac{R_m}{R}-\frac{R}{R_d}\right]\textrm{sech}^2(z/2z_d).
\end{equation}
For the $H_1$ disk, $z_d=0.085\textrm{ kpc}$, $R_m=4\textrm{ kpc}$, $\Sigma_{0,d}=53.1 M_\odot/\textrm{pc}^2$, and $R_d=7\textrm{ kpc}$. For the $H_2$ disk, $z_d=0.045\textrm{ kpc}$, $R_m=12\textrm{ kpc}$, $\Sigma_{0,d}=2180 M_\odot/\textrm{pc}^2$, and $R_d=1.5\textrm{ kpc}$.

The dark matter halo of \citeauthor{McMillan} has the same functional form as that of \citeauthor{Sofue}, but with $\rho_{0,h}=0.00854 M_\odot/\textrm{pc}^3$ and $h=19.6\textrm{ kpc}$. These will be the values parameterized in order to test the \citeauthor{McMillan} model.

\subsubsection{Piffl Model}
The model presented by \citet{Piffl} utilizes five components: a gas disk, a thin disk, a thick disk, a bulge, and a component for the dark matter halo. Each disk is described using
\begin{equation}
\rho(R,z)=\frac{\Sigma_0}{2z_d}\exp\left[-\left(\frac{R}{R_d}+\frac{|z|}{z_d}+\frac{R_{\textrm{hole}}}{R}\right)\right].
\end{equation}
In their work, \citeauthor{Piffl} assert some parameters for the gas disk, the bulge, and the dark matter halo, while other parameters for the thin, thick, and gas disks, and for the dark matter halo, are fit to observations. For the gas disk, $\Sigma_0=94.5 M_\odot/\textrm{pc}^2$,$R_d=5.36\textrm{ kpc}$, $z_d=0.04\textrm{ kpc}$, and $R_{\textrm{hole}}=4\textrm{ kpc}$.  For the thin disk, $\Sigma_0=570.7 M_\odot/\textrm{pc}^2$,$R_d=2.68\textrm{ kpc}$, $z_d=0.20\textrm{ kpc}$, and $R_{\textrm{hole}}=0\textrm{ kpc}$. For the thick disk, $\Sigma_0=251.0 M_\odot/\textrm{pc}^2$,$R_d=2.68\textrm{ kpc}$, $z_d=0.70\textrm{ kpc}$, and $R_{\textrm{hole}}=0\textrm{ kpc}$.

The bulge and dark matter halo are described using the same function:
\begin{equation}
\rho(R,z)=\frac{\rho_0}{m^\gamma(1+m)^{\beta-\gamma}}\exp[-(mr_0/r_\textrm{cut})^2],
\end{equation}
with $m(R,z)=\sqrt{(R/r_0)^2+(z/qr_0)^2}$. For the bulge, $\rho_0=9.49\times 10^{10} M_\odot/\textrm{kpc}^3$, $q=0.5$, $\gamma=0$, $\beta=1.8$, $r_0=0.075\textrm{ kpc}$, and $r_{cut}=2.1\textrm{ kpc}$. For the dark matter halo, $\rho_0=0.01816 M_\odot/\textrm{kpc}^3$, $q=1$, $\gamma=1$, $\beta=3$, $r_0=14.4\textrm{ kpc}$, and $r_{cut}=10^5\textrm{ kpc}$. We seek to constrain $\rho_0$ and $r_0$ for the dark matter halo.

\begin{figure}
	\includegraphics[width=\linewidth]{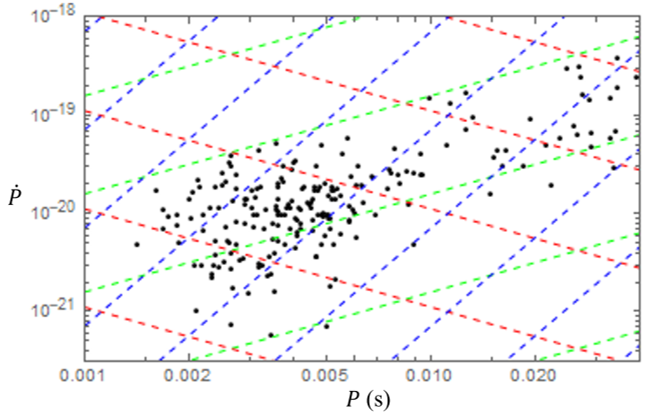}
	\caption{$P$-$\dot{P}$ diagram of our sample. Lines of constant spin-down luminosity (i.e., the rate of rotational kinetic energy loss) are shown in blue.  Lines of constant magnetic field strength are shown in red. Lines of constant inferred age assuming $\tau=P/2\dot{P}$ are shown in green.}
		\label{fig:1}
\end{figure}

\section{Data and Application of Method}
\subsection{Data Set}
Owing to their capacity for high-precision timing ($\dot{P}/P\sim10^{-18}$), we limited our data set to the use of millisecond pulsars (MSPs). Our MSP data set was obtained from \citet{ATNF}.  The method applied in this work requires known astrometric properties, such as position, distance, and proper motion, as well as pulsar timing properties (i.e., the first and second derivatives of frequency, $\dot{\nu}$ and $\ddot{\nu}$).  Of particular interest are PSR J1012+5307, PSR J1903+0327, and PSR J1959+2048, for which, in addition to all of the aforementioned quantities, the radial velocity is also known from \citet{Callanan}, \citet{Khargharia}, and \citet{vanKerkwijk}, respectively.  In these cases, it is possible to determine the contribution to $\ddot{\nu}$ directly rather than inferring the radial velocity from it.

Figure \ref{fig:1} shows a $P-\dot{P}$ diagram of pulsars used.  This set of pulsars was selected for the availability of frequency and frequency derivative information, distance and angular position, transverse velocities, proximity to the solar system, and age of the pulsar.  Frequency, frequency derivatives, distance, angular position, and transverse velocity are all necessary for the constraint of Galactic potential parameters.  Pulsars closer to the Galactic center must outshine background radiation, and are therefore thought to have stronger magnetic fields.  To reduce bias toward high-magnetic field pulsars, nearby pulsars were selected.  Younger pulsars, such as the Crab pulsar, are not suitable for constraint of parameters due to their rotational instability and were also removed.  This is accomplished by removal of pulsars with characteristic ages $\tau=P/2\dot{P}<10^9 \textrm{ yrs}$ and constraint of our dataset to pulsars with Galactic orbital radius $R > 5 \textrm{ kpc}$.  Additionally, pulsars in globular clusters were excluded in order to mitigate further ephemeral effects.

\subsection{Halo Mass}
\begin{figure}
	\includegraphics[width=\linewidth]{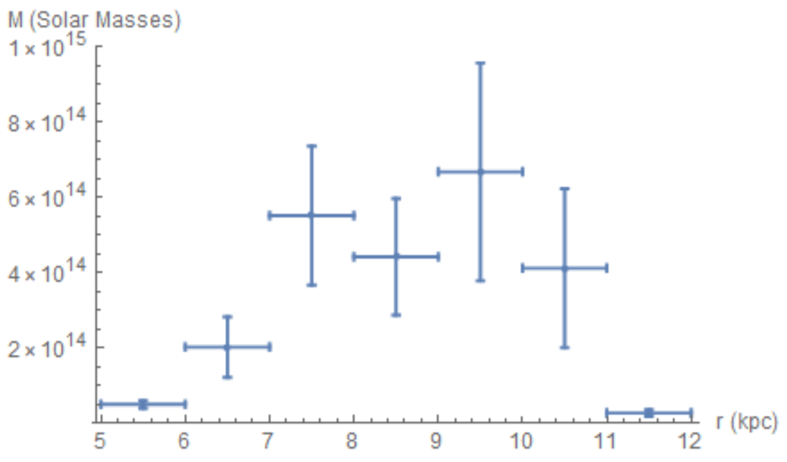}
	\caption{Constrained halo mass, given in solar masses, as inferred from pulsar spin period derivatives at various distances, given in kpc. Pulsars were binned into 1 kpc intervals.  Error bars were produced by bin size for radius and standard deviation divided by the square root of the number of data points in each bin for inferred masses.}
		\label{fig:2}
\end{figure}

\begin{figure}
	\includegraphics[width=\linewidth]{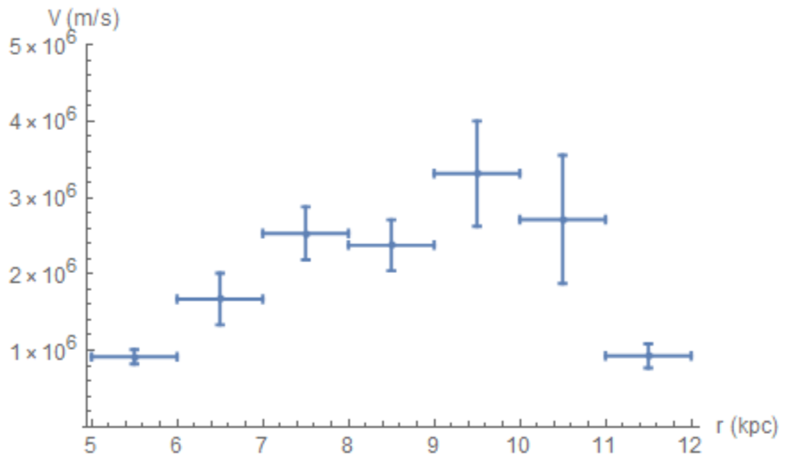}
	\caption{Virialized velocities, given in m/s, as inferred from pulsar spin period derivatives at various distances, given in kpc. Pulsars were binned into 1 kpc intervals.  Error bars were produced by bin size for radius and standard deviation divided by the square root of the number of data points in each bin for virialized velocities.}
		\label{fig:3}
\end{figure}

With transverse and radial velocities, positions, distances, $\nu$, and $\dot{\nu}$ in hand, we can use the approach outlined in \S 2 to infer $M_h$.  The halo mass in the \citeauthor{Kenyon} model inferred from spin period as a function of pulsar orbital radius is shown in Figure \ref{fig:2}.  From here, it is possible to calculate virial velocities corresponding to those of the flat rotation curves commonly used to infer dark matter.  These values are shown in Figure \ref{fig:3}.

\begin{figure}
	\includegraphics[width=\linewidth]
{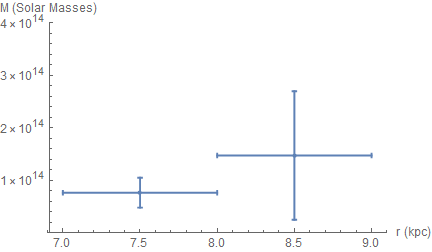}
	\caption{Constrained halo mass, given in solar masses, as inferred from pulsar binary orbital period derivatives at various distances, given in kpc. Pulsars were binned into 1 kpc intervals.  Error bars were produced by bin size for radius and standard deviation divided by the square root of the number of data points in each bin for inferred masses.}
		\label{fig:4}
\end{figure}

\begin{figure}
	\includegraphics[width=\linewidth]{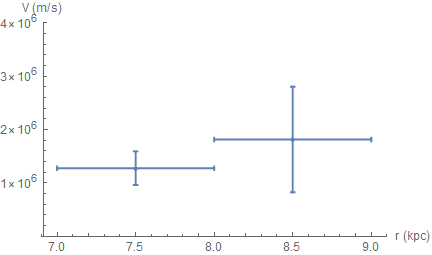}
	\caption{Virialized velocities, given in m/s, as inferred from pulsar binary orbital period derivatives at various distances, given in kpc. Pulsars were binned into 1 kpc intervals.  Error bars were produced by bin size for radius and standard deviation divided by the square root of the number of data points in each bin for virialized velocities.}
		\label{fig:5}
\end{figure}

While the magnitudes of halo mass and virialized velocity are somewhat high, one must bear in mind that magnetic spin-down has not yet been accounted.  As was previously stated, this problem can, in principle, be avoided by using binary orbital periods to infer halo mass. This is done in Figure \ref{fig:4}, with the corresponding virial velocities shown in Figure \ref{fig:5}. Data for binary orbital period derivatives were obtained from \citet{Zhu} and \citet{Matthews}.  Ideally, Figures \ref{fig:3} and \ref{fig:5} would reproduce the flat rotation curves commonly used to infer the existence of dark matter, but even in the case of binary orbital periods, timing is insufficiently constrained to produce a physically realistic result.  Nevertheless, if the ephemeral effects on the binary orbital period do not vary wildly with distance from the Galactic center, we should expect the true velocity curve to match the shape of Figure \ref{fig:5}, if not the magnitude.

\subsection{Magnetic Field Strength}
The magnetic spin-down due to a loss in rotational kinetic energy through magnetic dipole radiation is described by \citep{Longair}
\begin{equation}\label{B Pdot}
B=\sqrt{\frac{3\mu_0c^3M}{80\pi^3R^4}}(P\dot{P})^{1/2}\approx(3\times10^{15} \textrm{ T s}^{-1/2})(P\dot{P})^{1/2}\equiv c_B(P\dot{P})^{1/2}.
\end{equation}
For MSPs with the appropriate observations, it is possible to infer magnetic field strengths from $\ddot{\nu}$ as well, assuming that the magnetic field strength, the density, and the radius of the MSP are stable (i.e., magnetic spin-down only affects rotational frequency). The effect of magnetic spin-down is
\begin{equation}\label{B nudot}
\frac{\dot{\nu}}{\nu}=-\frac{B^2\nu^2}{c_B^2},
\end{equation}
which yields
\begin{equation}\label{ddotnu B}
B=3^{-1/4}\frac{c_B}{\nu}\left(\frac{\ddot{\nu}}{\nu}\right)^{1/4}\approx0.7598\frac{c_B}{\nu}\left(\frac{\ddot{\nu}}{\nu}\right)^{1/4}.
\end{equation}
In principle, it is possible to use $\dot{\nu}$ and $\ddot{\nu}$ observations in conjunction to find a better constraint for halo mass, but at present we find that uncertainties in $\ddot{\nu}$ timing are too great for such a method to be viable.  

Alternatively, we can use the values for magnetic field strength obtained from \eqref{B Pdot} to estimate the effect on $\ddot{\nu}$ assuming \eqref{ddotnu B}. Hence
\begin{equation}
\frac{\ddot{\nu}}{\nu}=3\left(\frac{B\nu}{c_B}\right)^4.
\end{equation}
The constraint of the magnetic field using observed $\ddot{\nu}$ was attempted, but found to be physically unrealistic due to poorly corrected $\ddot{\nu}$ values.  Reduction of $\ddot{\nu}$ residues to $\sim10^{-30}$ might yield better results.

\subsection{Radial Velocity Contributions}
If we allow the kinematic effects to comprise the entire observed $\ddot{\nu}$, then \eqref{Kinematic Contribution} provides an estimate of the radial velocity, subject to the systematic error introduced by intrinsic spin-down effects. For many pulsars, the constrained radial velocity was also found to be physically unrealistic, and was often in excess of $c$.  Nevertheless, future improvements to pulsar timing might make \eqref{Kinematic Contribution} a feasible method for ascertaining the radial velocity of pulsars which lack a binary companion with a well-understood spectrum. 

For PSR J1012+5307, PSR J1903+0327, and PSR J1959+2048, the radial velocities have been observed through spectroscopic measurements of binary companions. While the observed values, $v_r=44\pm8 \textrm{ km s}^{-1}$, $v_r=42.1\pm2.5 \textrm{ km s}^{-1}$, and $v_r=85\pm5 \textrm{ km s}^{-1}$, respectively, corroborate our limits, the velocities of the limiting cases are so large that they do not provide new information.  This again demonstrates the need for improvements in observations of $\ddot{\nu}$.

For the above pulsars with known radial velocities, we can calculate the kinematic contribution to $\ddot{\nu}$ directly using Equation \eqref{Kinematic Contribution}. For PSR J1012+5307, PSR J1903+0327, and PSR J1959+2048, \eqref{Kinematic Contribution} yields $\ddot{\nu}_{\textrm{kin}}=-4.20\times10^{-34} \textrm{ s}^{-3}$, $\ddot{\nu}_{\textrm{kin}}=-3.76\times10^{-34} \textrm{ s}^{-3}$, and $\ddot{\nu}_{\textrm{kin}}=-3.10\times10^{-34} \textrm{ s}^{-3}$, respectively. With observed $\ddot{\nu}\sim10^{-28}$, the kinematic contribution is clearly negligible.

\subsection{Constraint of Galactic Models Using PSR J1713+0747}
\citeauthor{Zhu} provide values for orbital $\nu$ and $\dot{\nu}$ of PSR J1713+0747, which have been corrected for ephemeral effects, along with an evaluation of the Shklovskii effect and the expected kinematic contribution to $\dot{\nu}$.  The values obtained therein are found to be in agreement with our own using values of position and proper motion from \citeauthor{ATNF}. However, as before, we seek to remove the assumption of a flat rotation curve.  This will allow us to constrain parameters for the various Galactic models mentioned in \S 2.

\begin{figure}
	\includegraphics[width=\linewidth]{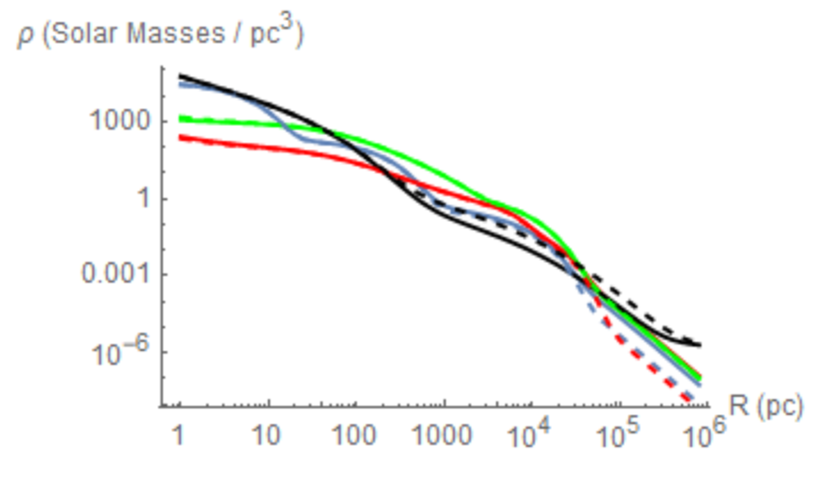}
	\caption{Galactic mass density $\rho$ as a function of Galactic radius in the Galactic plane.  Solid lines denote uncorrected values, i.e., those reported by each model's source, while dashed lines denote values modified to eliminate the timing residue. The blue, red, green, and black curves represent the Sofue, McMillan, Piffl, and Kenyon models, respectively.}
		\label{fig:6}
\end{figure}

The Shklovskii effect on orbital $\dot{\nu}/\nu$ is found to be $-1.056 \times 10^{-19} \textrm{ s}^{-1}$, such that $\dot{P}_b=6.18\times10^{-13}$. The residual $\dot{\nu}/\nu$ after correcting for the Shklovskii effect is $4.75\times10^{-20}\textrm{ s}^{-1}$, with $\dot{P}_b=-2.78305\times10^{-13}$.  Just as before, we use this residue to constrain parameters describing the Galactic halo. Some of the Galactic models used are summarized concisely in \citet{Perera}, whose Figure 3 is found to be in agreement with the uncorrected distributions in our Figure \ref{fig:6}.

After constraining the Galactic models above, we found that the following values for each model's parameters were better suited for explaining the $\dot{P}_b$ residue:

For the Sofue model, $\rho_{0,h}=0.01885^{+0.0058}_{-0.0058} M_\odot/\textrm{pc}^3$ and $h=6.5^{+2.0}_{-2.0}\textrm{ kpc}$ produce a minimum residue of $\dot{P}_b/P_b=2.94\times 10^{-20} \textrm{ s}^{-1}$, with a confidence interval of $7.5\%$.

For the McMillan model, $\rho_{0,h}=0.026047^{+0.001708}_{-0.026047} M_\odot/\textrm{pc}^3$ and $h=4.90^{+3.92}_{-4.90}\textrm{ kpc}$ produce a minimum residue of $\dot{P}_b/P_b=1.90\times 10^{-20} \textrm{ s}^{-1}$, with a confidence interval of $5\%$.

For the Piffl model, $\rho_{0,dm}=0.062652^{+0.007264}_{-0.058112} M_\odot/\textrm{pc}^3$ and $r_{0,dm}=9.4^{+31.7}_{-9.4}\textrm{ kpc}$ produce a minimum residue of $\dot{P}_b/P_b=2.35\times 10^{-20} \textrm{ s}^{-1}$, with a confidence interval of $17.5\%$.

For the Kenyon model, $M_h=3.25^{+0.60}_{-2.60}\times 10^{12} M_\odot$ and $r_h=21^{+24}_{-12}\textrm{ kpc}$ produce a minimum residue of $\dot{P}_b/P_b=1.91\times 10^{-22} \textrm{ s}^{-1}$, with a confidence interval of $98\%$.

\begin{figure}
	\includegraphics[width=\linewidth]{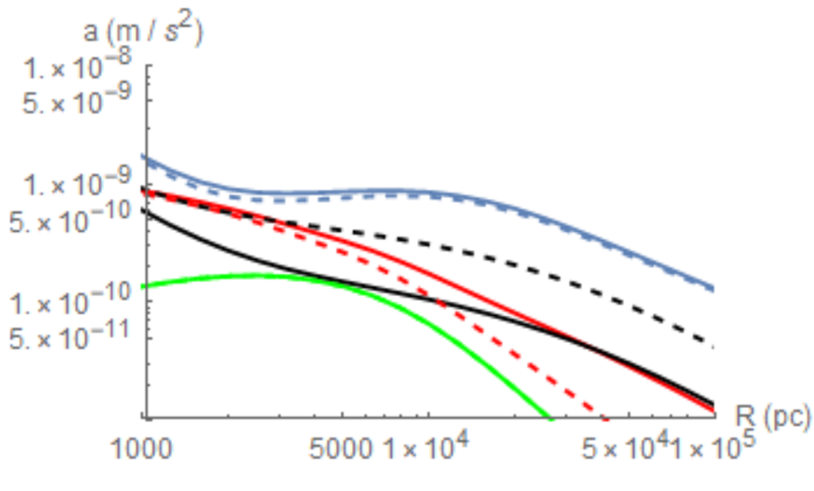}
	\caption{Acceleration $a$ as a function of Galactic radius in the Galactic plane.  Solid lines denote uncorrected values, i.e., those reported by each model's source, while dashed lines denote values modified to eliminate the timing residue. The blue, red, green, and black curves represent the Sofue, McMillan, Piffl, and Kenyon models, respectively.}
		\label{fig:7}
\end{figure}

The mass distributions and accelerations produced by these parameters can be seen in the dashed lines of Figures \ref{fig:6} and \ref{fig:7}.  Error bars in model parameters were determined by applying a $\chi^2$ statistic to the parameter values tested. The $\chi^2$ statistic itself was calculated from numerical solution of
\begin{equation}
p(\nu,\chi^2)=\frac{1}{\Gamma(\nu/2)}\int^\infty_{\chi^2/2}t^{\frac{\nu}{2}-1}e^{-t}dt,
\end{equation}
where $p(\nu,\chi^2)$ is the $p$-value, $\nu$ is the number of degrees of freedom (in this case $\nu=1$), and $\Gamma(x)$ is the gamma function.
The confidence interval was lowered until the parameter bounds fell within tested parameters. For the Kenyon model, numerical solution of the Poisson equation was not necessary, drastically decreasing computation time.  Since computation time is currently the limiting factor of this approach, this resulted in the $98\%$ confidence interval mentioned.

\section{Conclusion}

We have seen that observations of $\nu$, $\dot{\nu}$, distance, position, and transverse velocity are sufficient for an estimate of Galactic acceleration.  After replacing the contribution to $\dot{\nu}$ due to acceleration as calculated under the assumption of virialized velocities, we produce an estimate for Galactic halo mass by assuming that the resultant timing residue was due entirely to the kinematic effect of motion about the Galactic center.  

We found reasonable agreement with expected values for each model, with, e.g., the Kenyon halo mass $M_h=3.25^{+0.60}_{-2.60}\times 10^{12} M_\odot$.  Comparing this with $M_{h}=10^{12} M_{\odot}$, proposed by \citeauthor{Kenyon}, we find consistency between the direct measurement and conventional views about dark matter involving flat rotation curves. Furthermore, the halo radius $r_h=21^{+24}_{-12}\textrm{ kpc}$ is in agreement with the expected $r_h = 20\textrm{ kpc}$ from \citeauthor{Kenyon}, and with a residue of only $\dot{P}_b/P_b=1.91\times 10^{-22} \textrm{ s}^{-1}$ as a result.

For the \citeauthor{Sofue} model, we found $\rho_{0,h}=0.01885^{+0.0058}_{-0.0058} M_\odot/\textrm{pc}^3$, which is significantly smaller than the expected $\rho_{0,h}=0.029 M_\odot/\textrm{pc}^3$, along with $h=6.5^{+2.0}_{-2.0}\textrm{ kpc}$, compared to the expected $h=10\textrm{ kpc}$.  This produced a residue of $\dot{P}_b/P_b=2.94\times 10^{-20} \textrm{ s}^{-1}$.

With regards to the \citeauthor{McMillan} model, we obtained $\rho_{0,h}=0.026047^{+0.001708}_{-0.026047} M_\odot/\textrm{pc}^3$ and $h=4.90^{+3.92}_{-4.90}\textrm{ kpc}$, compared to the expected values of $\rho_{0,h}=0.00854 M_\odot/\textrm{pc}^3$ and $h=19.6\textrm{ kpc}$, with a residue of $\dot{P}_b/P_b=1.90\times 10^{-20} \textrm{ s}^{-1}$.

In the \citeauthor{Piffl} model, we found $\rho_{0,dm}=0.062652^{+0.007264}_{-0.058112} M_\odot/\textrm{pc}^3$ and $r_{0,dm}=9.4^{+31.7}_{-9.4}\textrm{ kpc}$, compared to the expected $\rho_{0,dm}=0.01816 M_\odot/\textrm{pc}^3$ and $r_{0,dm}=14.4\textrm{ kpc}$, with a residue of $\dot{P}_b/P_b=2.35\times 10^{-20} \textrm{ s}^{-1}$.

Further reduction of the pulsar residue is possible in each of these cases, but, with the exception of the \citeauthor{Kenyon} model, this approach was computationally limited due to the necessity of solving Poisson's equation numerically.  With greater computation time, more meaningful constraint of the parameters in each Galactic model is an immediate possibility even with current pulsar data.  Regardless, each model was brought to closer agreement in that the pulsar residue was reduced in this work.

If we assume the timing residual reported by \citeauthor{ATNF} is attributed to magnetic spin-down effects, we obtain an estimate for the magnetic field of each pulsar. We found that this value is typically $B\sim10^{8}$ Gauss. Since we believe the magnetic and Galactic effects to be the dominant contributors to timing residuals in $\dot{\nu}$, we sought to constrain the magnetic field using $\ddot{\nu}$ measurements in order to obtain more precise $\dot{\nu}$ values (and therefore more precise values for halo mass).  Unfortunately, due to the lack of precision in $\ddot{\nu}$ observations and a lack of data involving second-derivative orbital effects, this approach was ineffective. Data obtained from such an approach was therefore subject to systematic error caused by magnetic spin-down effects.

Considering the functional form of the Galactic effect on $\ddot{\nu}$, we noted its dependence on radial velocity.  Assuming that the $\ddot{\nu}$ timing residual was entirely due to kinematic effects, we were able to obtain limits on the radial velocity of each pulsar.  We encountered the same issue as before, i.e., that pulsar timing in $\ddot{\nu}$ remains insufficient for constraint of MSP properties. Nevertheless, we have here produced a method for such constraint should better pulsar data become available. 

Clearly, much of the aforementioned work will be facilitated by improvements to $\ddot{\nu}$ data.  To obtain such data, it is necessary to further constrain various effects which dominate the $\ddot{\nu}$ residual.  This, in turn, requires the characterization of binary orbital, line-of-sight, and magnetic spin-down $\ddot{\nu}$ contributions. If, in addition to these, more direct estimates of pulsar magnetic field strengths become available, it would be possible to fully account for the magnetic dipole radiation contribution to $\ddot{\nu}$.  While improvements of $\ddot{\nu}$ timing to the necessary values of $\ddot{\nu}\sim10^{-30}$ may seem daunting, this still seems plausible in light of drastic improvements to $\dot{\nu}$ timing since the discovery of the Hulse-Taylor binary.  We note that similar efforts which enlist pulsar timing as a means of accelerometry are already underway viz. \citet{Chakrabarti} and \citet{Phillips}.

\begin{table}
\begin{tabular}{|l|l|l|l|}
\hline
\textbf{Model} & \textbf{Characteristic $\rho$ or $M$ ($M_\odot$)}                    & \textbf{Radius (kpc)}      & \textbf{Confidence} \\ \hline
Sofue          & $0.01885^{+0.0058}_{-0.0058} /\textrm{pc}^3$      & $6.5^{+2.0}_{-2.0}$    & $7.5\%$             \\ \hline
McMillan       & $0.026047^{+0.001708}_{-0.026047} /\textrm{pc}^3$ & $4.90^{+3.92}_{-4.90}$ & $5\%$               \\ \hline
Piffl          & $0.062652^{+0.007264}_{-0.058112} /\textrm{pc}^3$ & $9.4^{+31.7}_{-9.4}$   & $17.5\%$            \\ \hline
Kenyon         & $3.25^{+0.60}_{-2.60}\times 10^{12}$             & $21^{+24}_{-12}$       & $98\%$              \\ \hline
\end{tabular}
\caption{Characteristic density/mass, characteristic radius, and confidence interval for each model.}
\end{table}

\bibliographystyle{mnras}
\bibliography{bibliography}

\end{document}